\documentclass[aps,pre,final,groupedaddress,onecolumn,showpacs,showkeys,notitlepage ]{revtex4-1}
\usepackage{graphicx}
\usepackage{epstopdf}
\begin{document}

\title{Statistical distributions and entropy considerations in gene codes}


\author{Krystyna Lukierska-Walasek}
\email[]{k.lukierska@uksw.edu.pl}
\affiliation{ Faculty of Mathematics and Natural Sciences, Cardinal Stefan Wyszy\'nski University, W{\'o}ycickiego 1/3, 01-938 Warsaw, Poland }
\author{ Krzysztof Topolski}
\email[]{topolski@math.uni.wroc.pl}
\affiliation{Institute of Mathematics, Wroclaw University, Pl.  Grunwaldzki 2/4, 50-384 Wroc{\l}aw, Poland}
\author{ Krzysztof Trojanowski }
\email[]{trojanow@ipipan.waw.pl}
\affiliation{Institute of Computer Science, Polish Academy of Science, ul.Jana Kazmierza 5,01-248 Warsaw, Poland }
\date{\today}
\begin{abstract}
In our paper selected linguistic features of genomes to study the statistics of the gene codes are considered.
We present the information theory from which it follows that if the system is described by distributions of hyperbolic type it leads to the possibility of entropy loss and stability. We show that the histograms of gene lengths are similar to that of language words. We show the correspondence between presented theory and results for the number of replicated genes and replicated fragments of genes in genomes for {\em Borelia burgdorferi}, {\em Escherichia coli} and {\em Saccharomyces cerevisiae S288c}.
\end{abstract}
\pacs{87.85.mg, 87.10.Vg, 87.10.Mn}
\keywords{genome, gene code, gene length, statistical distributions, entropy}
\maketitle
\section{Introduction}
In last years there appeared a possibility to provide some knowledge of the genome sequence data in many organisms. The genome was studied intensively by number of different methods \cite{Holste2003,LiKa1994,Man1994,Man1995,Messer2005,Peng1992,Peng1994,Vieira1999,Stanley1999,Voss1992}. The statistical analysis of DNA is complicated because of its complex structure; it consists of coding and non coding regions, repetitive elements, etc., which have an influence on the local sequence decomposition. Long range correlations in sequence compositions of DNA are much more complex than simple power law; moreover effective exponents of scale regions are varying considerably between different species~\cite{Holste2003}, \cite{Messer2005}. In  papers~\cite{Man1994,Man1995} the Zipf approach to analyzing linguistic texts has been extended to statistical study of DNA base pair sequences.

In our paper we take into account some linguistic features of genomes to study the statistics of gene codes. One of the fundamental observations based on the information theory says that if the system is described by special distribution of hyperbolic type it implies a possibility of entropy loss. Distributions of hyperbolic type describe also property of stability and flexibility which explain that the languages can develop without changing its basic structure (see~\cite{HaTo2001}). In the experimental part of our research we show that a similar situation can occur in a genome sequence data which carries the genetic information.

To demonstrate analogies between coding of genes and representation of words in human languages we generated histograms of the sizes of groups of words of the same length in selected different European languages as well as histograms of the sizes of groups of genes of the same length in a genome. Three genomes were selected for evaluation, that is, two bacteria: {\em Borrelia burgdorferi} and {\em Escherischia coli}, and a yeast: {\em Saccharomyces cerevisiae S288c}. The gene data were obtained from NCBI GeneBank~\cite{GeneBank}. Each of the histograms was successfully approximated by Asymmetric Normal Inverse Gaussian Distribution.

Then, for each of the three genomes histograms of the sizes of groups of identical genes were generated. We showed that each of them can be modelled by distributions of hyperbolic type defined in the paper.

Additionally, the histograms for sizes of groups of identical fragments of genes of different sizes were calculated. In this case any two genes are regarded as similar when there exist a gene fragment, that is, an ordered sequence of symbols of size $n$ which can be found in both genes. Obtained results confirmed our hypothesis that the basic structure of genome remain stabile and entropy loss is possible.

The paper is organized as follows. In Sect.~\ref{commfeat} we begin with a brief presentation of some common features of linguistics and genetics. Then, in Sect.~\ref{MEPCLGEL} we describe Code Length Game, Maximal Entropy Principle and the theorem about hyperbolic type distributions and entropy loss. Sect.~\ref{Experim} contains the experimental research concerned analysis of statistical properties of gene strings in three selected genomes and consists of three parts. Histograms of gene lengths are presented in the first part.  Evaluation of the frequency of genes in a genome is the subject of the second  part, that is, the results of searching for replicated genes in three genomes are presented. In the third, last part we observe frequency of appearance of gene fragments in a genome for different lengths of the fragments. Sect.~\ref{Concl} concludes the paper.

\section{Some common features of linguistics and genetics}\label{commfeat}

A language is characterized by an alphabet with letters, like for the case of latin languages: \emph{a}, \emph{b}, \emph{c}, etc. The words are denoted as sequences of $n$ letters. In quantum physics the analogy to letters can be attached to pure states, and  texts correspond to mixed general states.  Languages can have very different alphabets, for example: computers --- 0,1 (two bits), English language --- 27 letters with space, or DNA --- four nitric bases: G({\em guanine}), A({\em adenine}), C({\em cytosine}), T({\em thymine}). The collection of letters can be ordered or disordered.
To quantify the disorder of different collections of the letters we use an entropy 
\begin{equation}
 \label{eq1}
  H = - \sum\limits_{i} p_{i}\log_{2}p_{i},
\end{equation}
where $p_{i}$ denotes probability of occurrence of the $i$-th letter.
If we take the base of logarithm $2$ this will lead to the entropy measured in bits. When all letters have the same probability in all states obviously the entropy has maximum value $H_{max}$.

A real entropy has lower value $ H_{eff}$, because in a real languages the letters have not the same probability of appearance. Redundancy $R$ of language is defined \cite{Shannon1948} as follows:
\begin{equation}
 \label{eq2}
  R = \frac{H_{max} - H_{eff}}{H_{max}}.
\end{equation}

The  quantity $H_{max}$\, - $H_{eff} $ is called an information.
Information depends on difference between the maximum entropy and the actual entropy. The bigger actual entropy  means the smaller redundancy.  Redundancy can be measured by values of the frequencies with which different letters occur in one or more texts.
Redundancy $R$ denotes the number that if we remove the part $R$ of the letters determined by redundancy, the meaning of the text will be still understood.

In English some letters occur more frequently than other and similarly in DNA of vertebrates the frequency of nitric bases C and G pairs is usually less than the frequency of A and T pairs.  The low value of redundancy allows in  easier way to fight transcription errors in a gene code. In the papers~\cite{Man1994,Man1995} it is demonstrated that non coding regions of eukaryotes display a smaller entropy and larger redundancy than coding regions.

\section{Maximal Entropy Principle and Code Length Game and Entropy Loss}\label{MEPCLGEL}
In this section we provide some mathematics from the information theory which will be helpful in the quantitative formulation of our approach, for details see~\cite{HaTo2001}.

Let $A$ be the $alphabet$ which is a discrete finite or countable infinite set.
Let $M_{+}^{1}$ and $^{\sim}M_{+}^{1}(A)$ are respectively, the set of probability measures on $A$ and the set of non-negative measures $P$, such that $P(A)\leq 1$.
The elements in $A$ can be thought as {\em letters}.
By $K(A)$ we denote the set of mappings,  {\em  compact codes}, $k\,:A \,\rightarrow[0 , \infty]$, which satisfy {\em Kraft's equality}~\cite{Cov1991}:
\begin{equation}
 \label{eq3}
  \sum\limits_{i\in A}\exp (-k_{i}) = 1.
\end{equation}
By $^{\sim}K(A)$ we denote the set of all mappings, {\em general codes}, $k\,: \, A\rightarrow[0 ,\infty]$, which satisfy  \emph{Kraft's  inequality}~\cite{Cov1991}:
\begin{equation}
 \label{eq4}
  \sum\limits_{i\in A}\exp (-k_{i}) \le 1.
\end{equation}
For $k\in ^{\sim}K(A)$ and $i\in A$,  $k_{i}$ is the {\em code length}, for example the length of the word.
For $k\in$ $^{\sim}K(A)$ and $P \in M_{+}^{1}(A)$ the {\em average code length}  is defined as
\begin{equation}
 \label{eq5}
 <k,P> =\sum\limits_ {i\in A} k_{i}p_{i}.
\end{equation}
There is bijective correspondence between $p_{i}$ and $ k_{i}$
$$ k_{i} = -\ln p_{i}   \quad \mbox{and} \quad p_{i}=\exp(-k_{i}).$$
For $P \in M_{+}^{1}(A)$ we also introduce the entropy
\begin{equation}
 \label{eq6}
 H(p) = -\sum\limits_{i\in A} p_{i}\ln p_{i}.
\end{equation}
The entropy can be represented as minimal average code length, (see \cite{HaTo2001}):
$$H(P)=\min_{k\in ^{\sim}K(A)}<k,P>.$$
Let ${\cal P}  \subseteq{ \cal M }_{+}^{1}(A)$  than
\begin{eqnarray}
\label{NashStrat}
 H_{\max}(\cal P)&=&\sup_{P\in \cal P} \inf_{k\in ^{\sim}K(A)}< k, P >\nonumber\\
&=&\sup_{P\subset \cal P}H(P)\nonumber\\
 &\leq&\inf_{k\in ^{\sim}K(A)} \sup_{P\subset \cal P}< k, P > \nonumber\\
&=& R_{min}(\cal P).
\end{eqnarray}
The formula~(\ref{NashStrat}) presents the Nash optimal strategies. ${\cal R}_{min}$ denotes $ minimum$ $risk$, $k$ denotes the Nash equilibrium code, $P$ denotes probability.  For example, in a linguistics the listener is a minimizer, speaker is a maximizer. We have words with distributions $p_{i}$ and their codes $k_{i}$, $i=1,2,... $. The listener chooses  codes $k_{i}$,the speaker chooses probability distributions $p_{i}$.\\
We can notice that Zipf argued~\cite{Zipf1949} that in the development of a language vocabulary balance is reached as a result of two opposing forces: $unification$ which tends to reduce the vocabulary and corresponds to a principle of least effort, seen from point of view of speaker and {\em diversification} connected with the listeners wish to know meaning of speech.

The principle~(\ref{NashStrat}) is so basic as {\em Maximum Entropy Principle} has a sense that search for one type of optimal strategy called as {\em  Code Length  Game} translates directly into a search for distributions with maximum entropy.
It is a given a code  $ k\in\,  ^{\sim}K(A) $ and distribution $P\in M^{1}_{+}(A)$. Optimal strategy according  $ H(P) = inf_{k\in\, ^{\sim}K}<k,P>$ is represented by entropy $H(P)$, where actual strategy is represented by $<k,P>$.

Zipf's law is an empirical observation which relates rank and frequency of words in natural language \cite{Zipf1949}. This law suggests modeling by distributions of hyperbolic type~\cite{HaTo2001}, because no distributions over  $N$ have probabilities proportional to $1/i$, due to the lack of normalization condition.\\ We consider a class of distributions $P= (p_{1},p_{2},. ..)$ over $N$.  If $ p_{1}\ge,p_{2}\ge ... $, $P$ is said to be hyperbolic if for any given $\alpha>1$,   $p_{i} \ge i^{-\alpha}$ for infinitely many indexes $i$.   As an example we can choose $ p_{i}\sim i^{-1}( \log(i) )^{-c}$ for some constant  $c>2$.

The Code Length Game for model ${\cal P}\in M_{+}^{1}(N)$  with codes  $k:  A \rightarrow  [0 ,\infty]$ for which $ \sum\limits_{i\in A}\exp (-k_{i}) = 1$, is in equilibrium if and only if $ H_{max}( co ({\cal P }) = H_{max}({\cal P})$. In such a case
a distribution $P^{*}$ is the $H_{max}$ attractor such that $P_{n}\rightarrow P^{*}$, if for every sequence $ (P_{n})_{n>1} \subseteq \cal P $  for which $H( P_{n}) \rightarrow H_{max}(\cal P)$. One expects that $ H(P^{*}) = H_{max}(\cal P)$ but in the case with entropy loss we have $ H(P^{*}) < H_{max}(\cal P)$.  Such possibility appears when distributions  are hyperbolic. It follows from  theorem  in \cite{HaTo2001}.\\

{\bf Theorem.} {\em
Assume that $P^{*}\in M_{+}^{1}(N)$  is of finite entropy and it has ordered point probabilities. The necessary and sufficient condition that $P^{*}$ can occur as $ H_{max}$--attractor in a model with entropy loss, is that $P^{*}$ is hyperbolic. If this condition is fulfilled than for every $h$ with
    $ H(P^{*})< h < \infty $, there exists a model ${\cal P} = {\cal P}_{h}$ with $P^{*}$ as    $H_{max}$--attractor and $ H_{max}({\cal P}_{h}) = h $.  In fact, ${\cal P}_{h} = ( P|<k^{*},P> \le h) $ is a largest model.  $k^{*}$ denotes the code adopted to $ P^{*}$, that is, $k^{*} = -\ln(p^{*})$, $ i>1$.}\\

As an example we can consider "an ideal" language where the frequencies of words are described by hyperbolic distribution $ P^{*}$ with a finite entropy. At a certain stage of life one is able to communicate at reasonably height rate about $H(P^{*})$ and improve language by introduction of specialized words, which occur seldom in a language as a whole. This process can be continued during the rest of life. One is able to express complicated ideas develop language without changing a basic structure of the language. We can expect similar situation in gene strings, which also carry an information.
\section{Application to genetics}\label{Experim}
\subsection{Part I-- Histograms for gene lengths and word lengths}
In this part we study the histograms of gene lengths. One can observe, that the histograms are of the same type as histograms for the word lengths. We model histograms of gene lengths by Asymmetric Inverse Gaussian Distributions as in the case of histograms of word lengths.
\begin{figure}[ht]
\centering
\includegraphics[width=5.5cm]{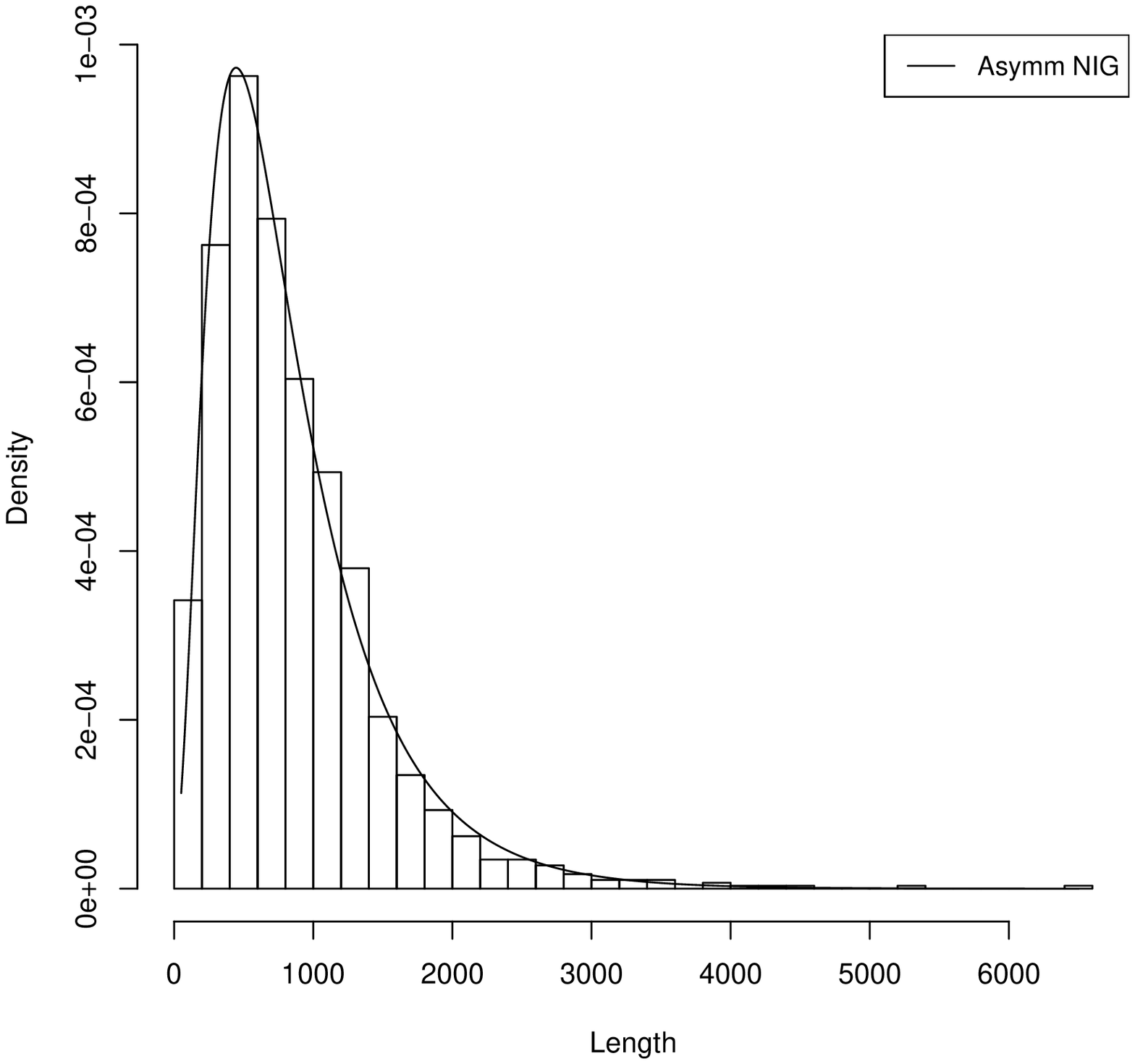}
\includegraphics[width=5.5cm]{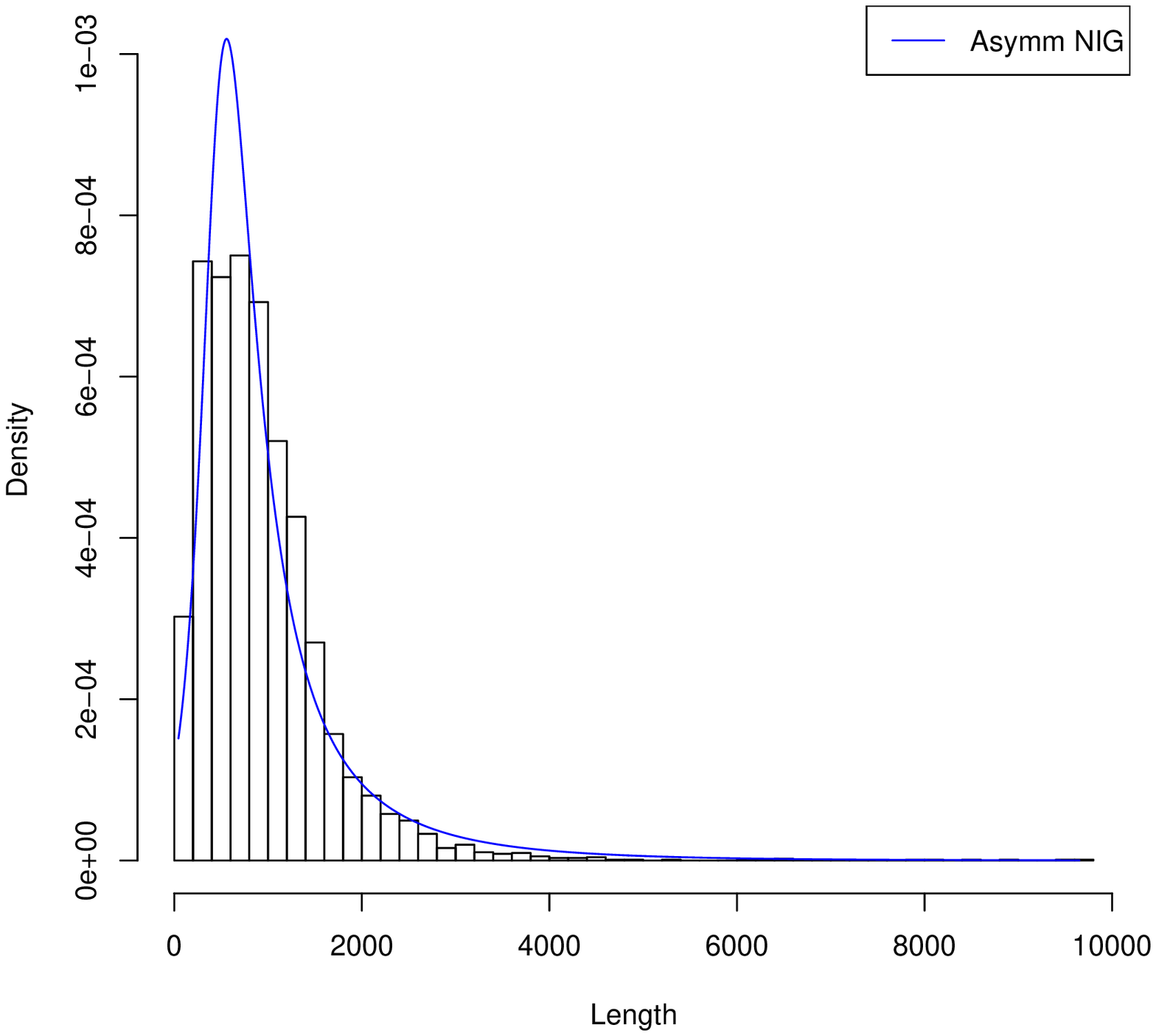}
\includegraphics[width=5.5cm]{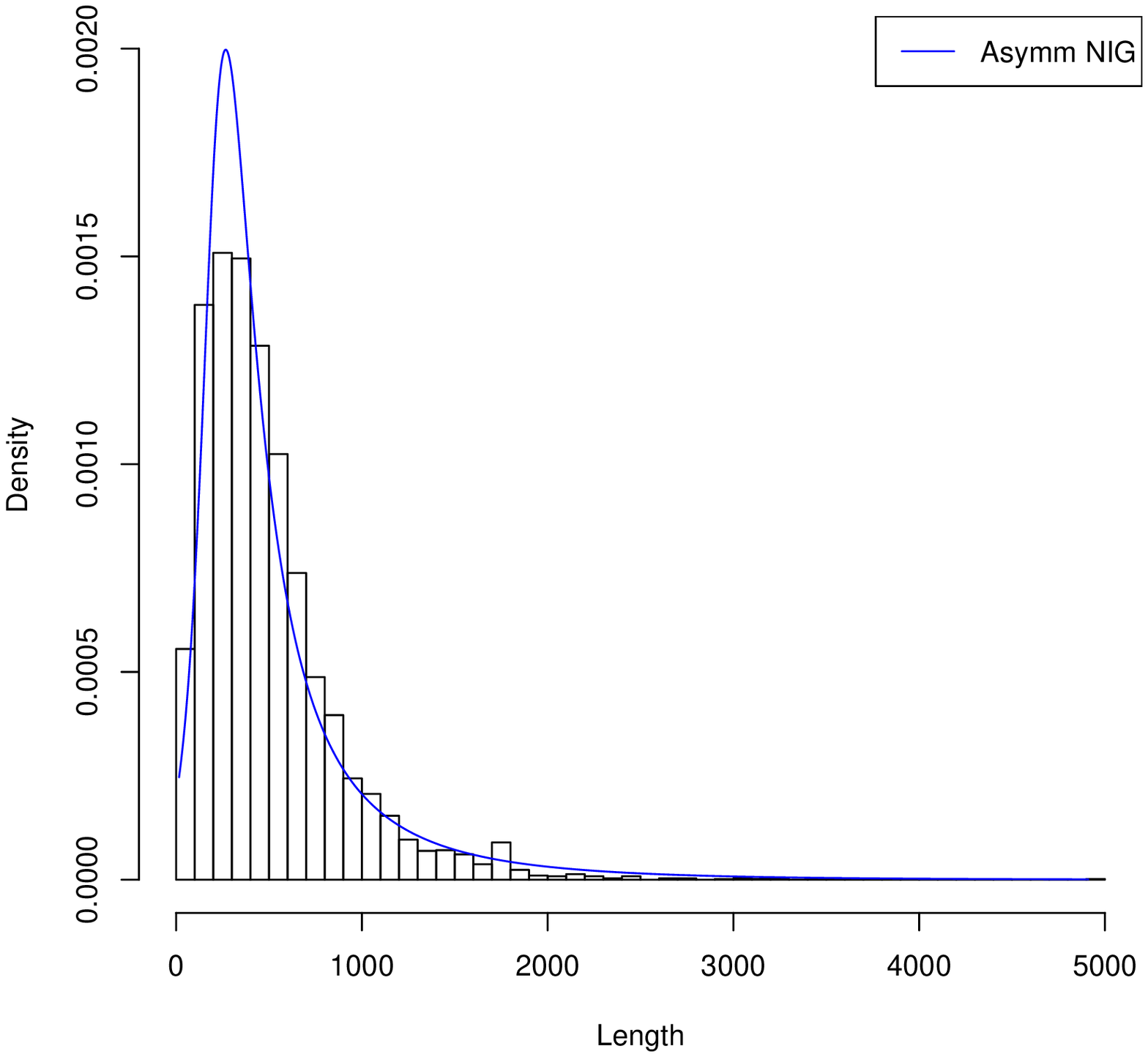}
\caption{Histograms of the gene length in the genome of {\em Borrelia burgdorferi} (the left graph), {\em Escherichia coli} (the middle graph) and {\em Saccharomyces cerevisiae S288c} (the right graph) with the fits of AIGD}\label{Fig1}
\end{figure}

Figure~\ref{Fig1} presents histograms of the gene length in three genome and their approximations by probability the Asymmetric Inverse Gaussian Distribution (AIGD). The AIGD has following parameters: for {\em Borrelia burgdorferi} ($\mu=420.9643$, $\sigma=466.0570$ and $\gamma=468.7277$); for {\em Escherischia coli} ($\mu=448.0338$, $\sigma=531.6735$ and $\gamma=529.9635$); for {\em Saccharomyces cerevisiae S288c} ($\mu=201.5467,$ $\sigma=257.1359$ and $\gamma=322.3161$).

\begin{figure}[ht]
\centering
\includegraphics[width=4.8cm]{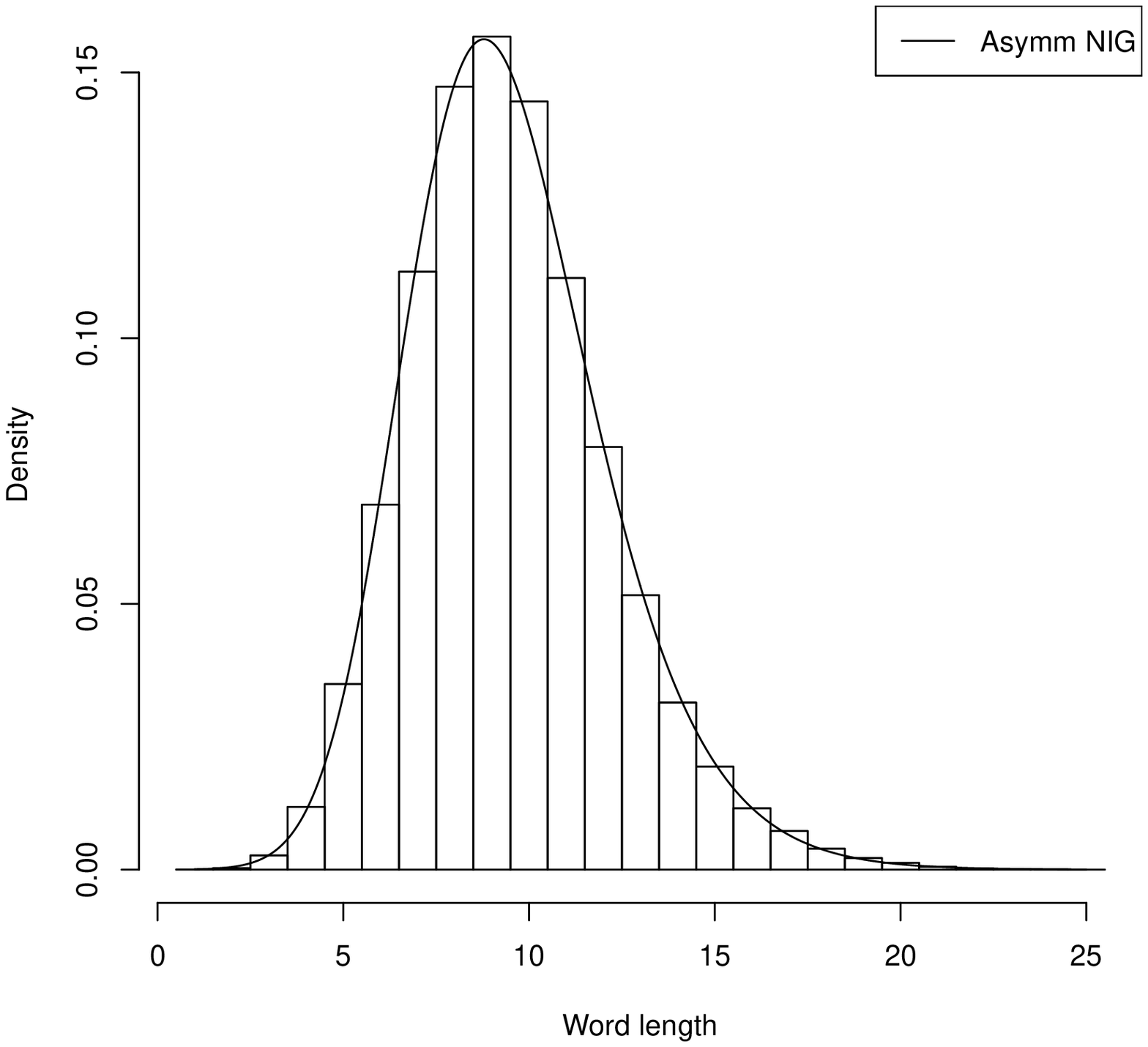}
\includegraphics[width=3.8cm]{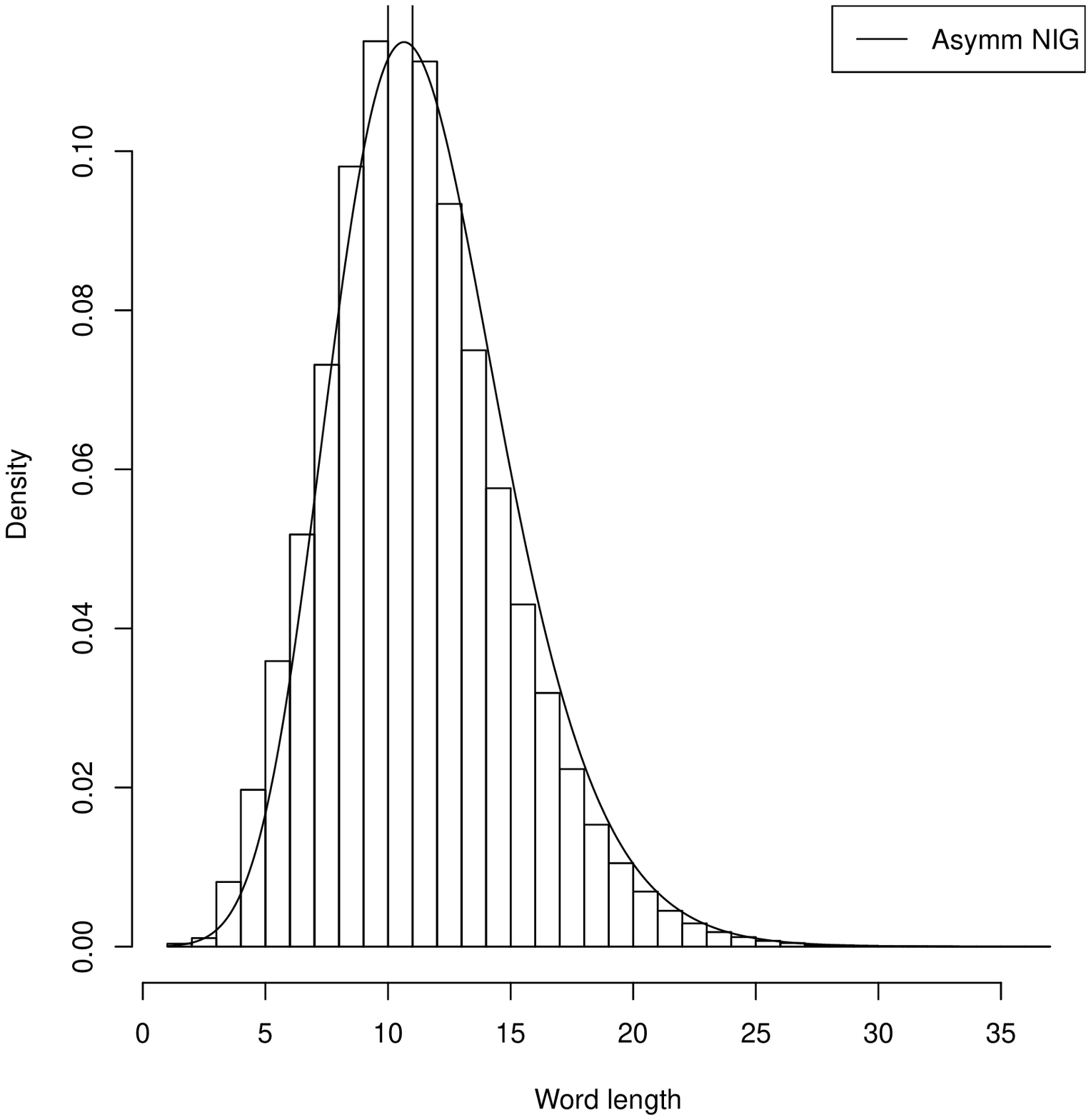}
\includegraphics[width=4.1cm]{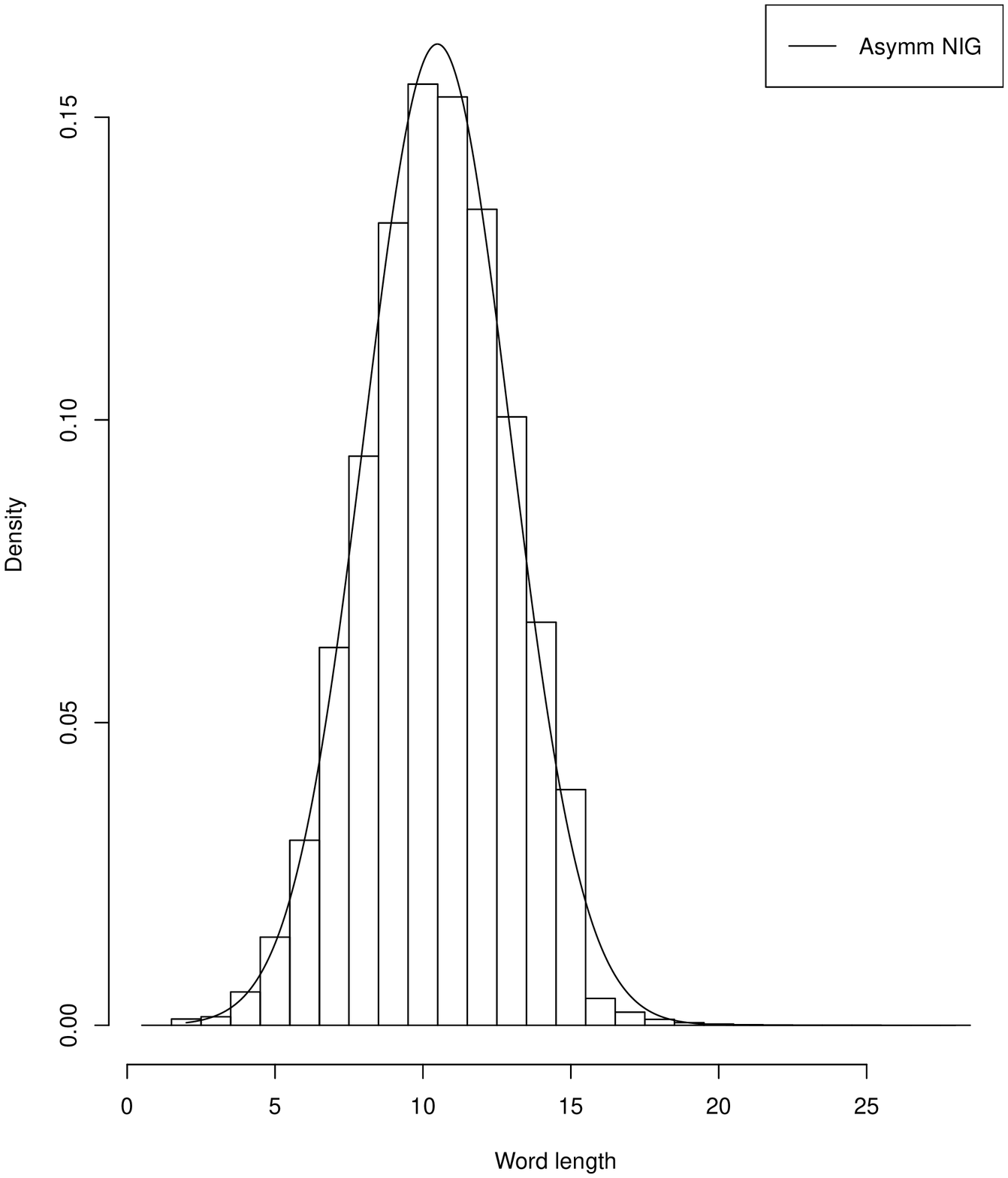}
\includegraphics[width=4.4cm]{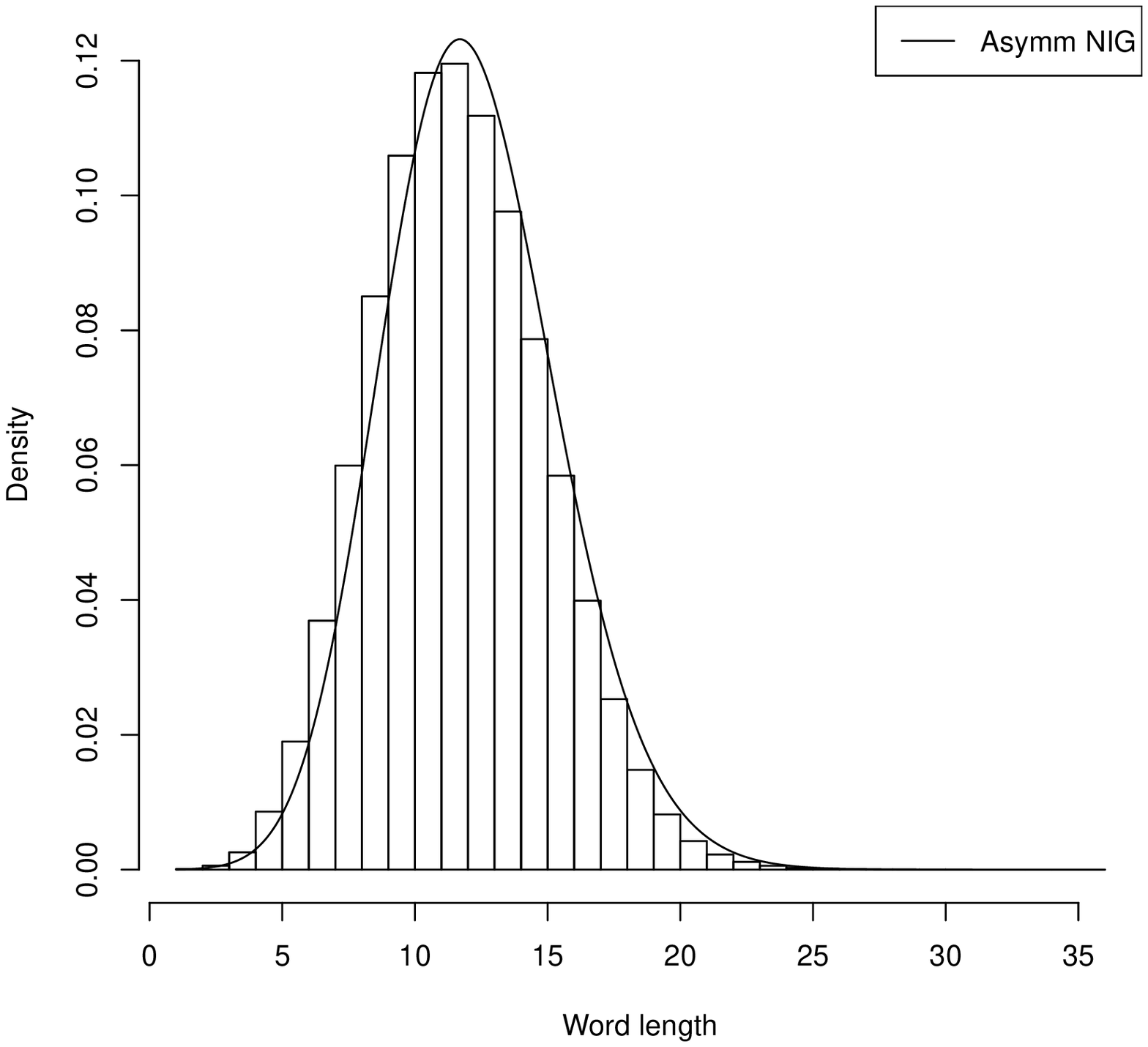}
\caption{\label{LangHist} Graphs of the sizes of groups of words of the same length in four different European languages from the left to the right, Czech, German, Italian and Polish respectively.}
\end{figure}

Systems where maximum entropy distributions does not exists can be described by distributions of hyperbolic type. In such systems stability and flexibility is present similarly as in natural languages or in description of genes where redundancy is confirmed.
It is interesting to show that the histograms for gene lengths resemble other histograms obtained for word lengths in human languages~\cite{WinEdt}. In Figure~\ref{LangHist} we present the histograms of the sizes of groups of words of the same length for four European languages: Czech, German, Italian and Polish which also the fit the Asymetric Inverse Gaussian distributions. 

 In this case the AIGD has following parameters:
for Czech ($\mu=4.0883,$ $\sigma=2.1168,$ and $\gamma=5.4246$);
for German ($\mu=-0.5941,$ $\sigma=2.2183,$ and $\gamma=12.2476$);
for Italian ($\mu=11.0250,$ $\sigma=2.4615$ and $\gamma=-0.5362$);
and for Polish ($\mu=-15.1809,$ $\sigma=1.3876,$ and $\gamma=27.3526$).
We investigated asymptotic properties of these distributions. The tails of these distributions appear to be of the power type. We show it for {\em Borrelia} and {\em Saccharomyces}.

\noindent

\begin{figure}[ht]
\centerline{\includegraphics[width=7cm]{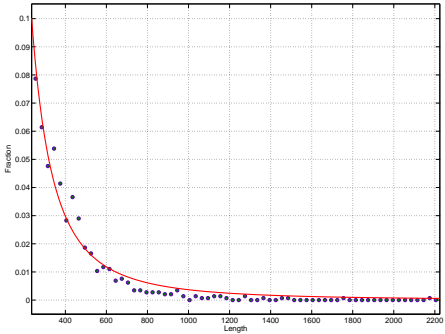}}
\caption{\label{Fig3} The comparison of the tail of gene length distribution of the {\em Borrelia burgdorferi} to the tail of the power type distribution.}
\end{figure}

Figure~\ref{Fig3} shows the fit of the tail of gene length distribution of the {\em Borrelia burgdorferi} to function of the form $f(x)=C\,x^{-1-a}$, where $C>0$ and $a>2$ are constants. The least squares best fit are obtained with $C=1500,899$ and $a=2,165$.

\begin{figure}[ht]
\centering
\includegraphics[width=6.5cm]{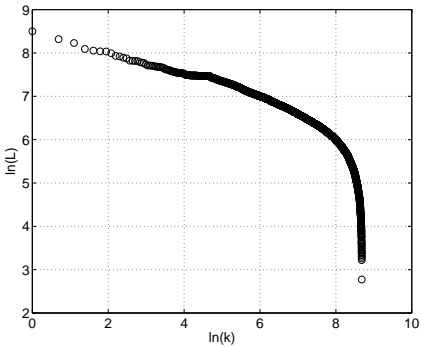}
\caption{\label{Fig5} Plot of the logarithm of the gene length $L$ versus logarithm of its length rank for the {\em Saccharomyces cerevisiae S288c} genome}
\end{figure}

Figure~\ref{Fig5} presents plot of the logarithm of gene length versus logarithm of its length rank for {\em Saccharomyces cerevisiae S288c}.

\begin{figure}[ht]
\centerline{\includegraphics[width=6.5cm]{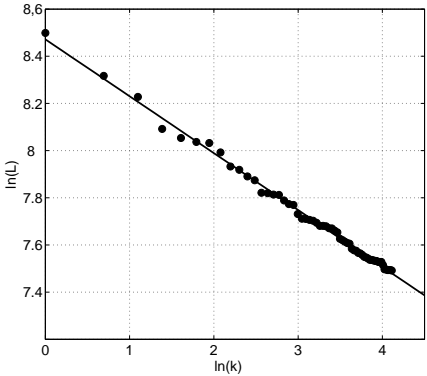}}
\caption{\label{Fig6} Plot of the logarithm of the gene length $L$ versus logarithm of it length rank for the genes with size bigger  then 5379 bp in  {\em Saccharomyces cerevisiae S288c} genome. The straight line represent the linear regression  line $y=-0.2413\,x+8.4718.$}
\end{figure}

Figure~\ref{Fig6} presents a part of the previous plot, the fit of the tail(for genes with size bigger than 5379 bp) of gene length distribution of the {\em Saccharomyces cerevisiae S288c}. The straight line denotes that tail of the gene length distributions is of power type.

\subsection{Part II -- Statistics for replicated genes}

In this part we evaluated numbers of identical genes in each of the three genomes. Any two genes are identical iff they are of the same length \emph{and} have the same content ordered in the same way. Otherwise, the genes are regarded as different, that is, for example, a smallest difference in the order of symbols in any two genes of the same size and content makes them different. The main parameters of the gene "language" for each on the genomes is presented in Table~\ref{table_params}. One can see, that gene lengths span within the range from 6381 for {\em Borrelia burgdorferi} to 14682 --- for {\em Saccharomyces cerevisiae S288c}. In every case the range is many times wider that the number of genes in the genome. However, in spite of this, some number of duplicated genes can be found in genomes of each of the analysed bacterium and yeast.
\begin{table}[ht]
\renewcommand{\arraystretch}{1.5}
\caption{Number of genes, maximum and minimum length of genes in the three selected genomes} \label{table_params}
\centering
{\smallskip}
{\smallskip}
{\smallskip}
{\smallskip}

\begin{tabular}{lccc}
\hline

{\small genome} & {\small num. of genes} & {\small min. length} & {\small max. length} \\
\hline
{\small \em Borrelia burgdorferi}                  & {\small 1242} &	{\small 120} &  {\small 6501} \\
{\small \em Escherischia coli}                     & {\small 4920} &	 {\small 75} & {\small 11421} \\
{\small \em Saccharomyces cerevisiae S288c}        & {\small 5906}  &	 {\small 51} & {\small 14733} \\
\hline

\end{tabular}
\end{table}

\begin{figure}[ht]
\centerline{
\includegraphics[width=6.2cm]{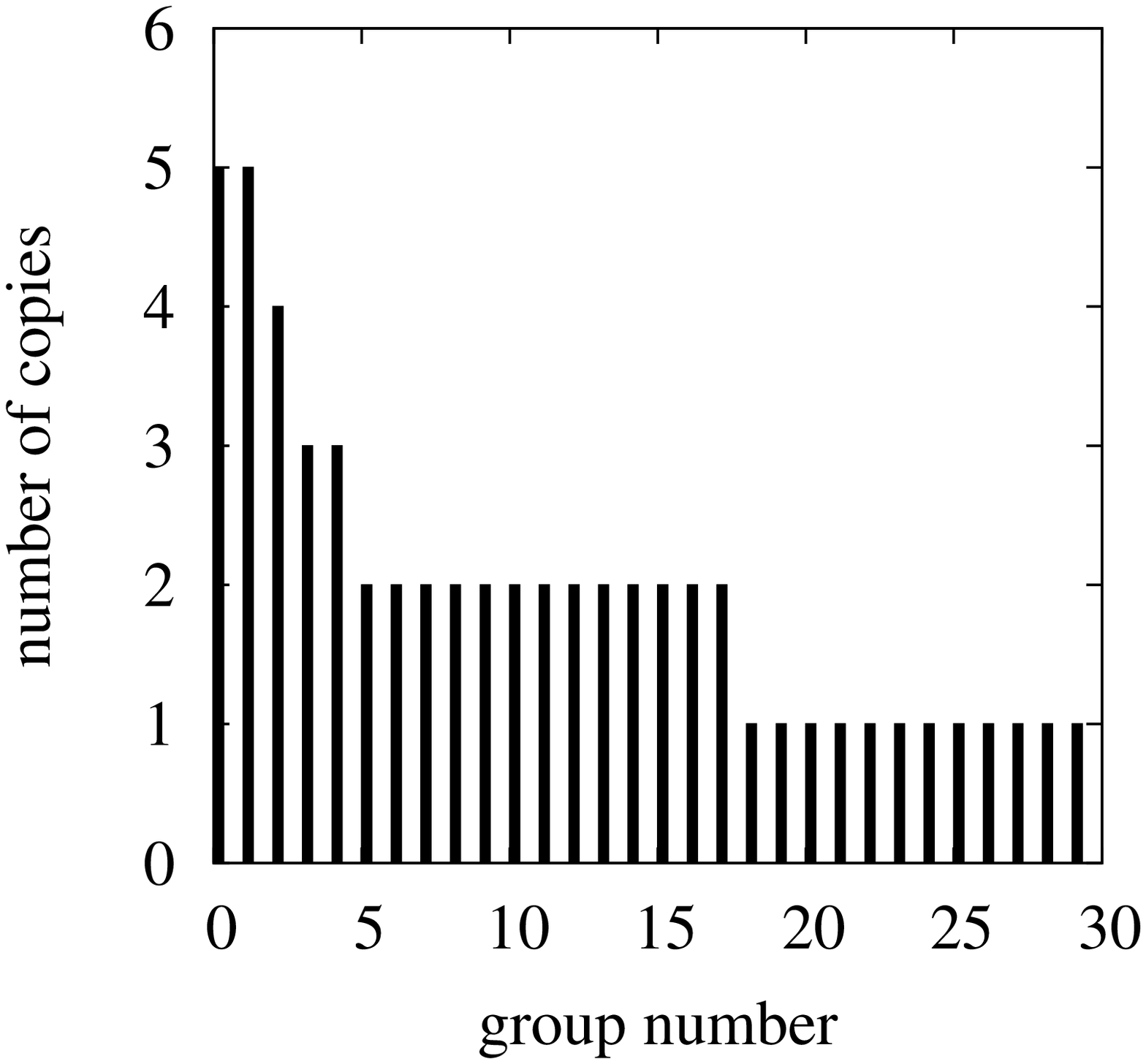}
\includegraphics[width=6.2cm]{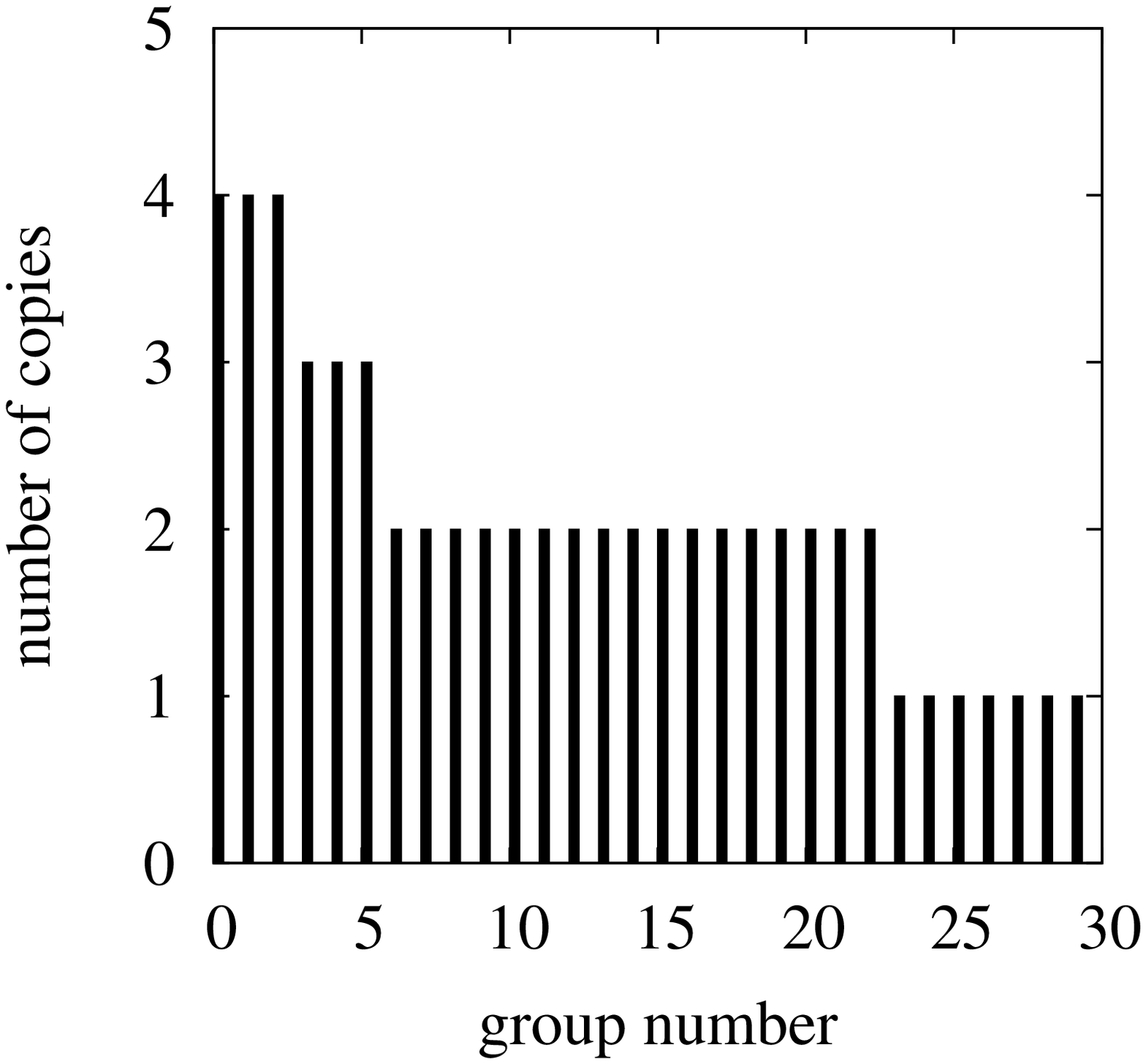}}
\caption{\label{FigHist} Graphs of the sizes of groups of identical genes in {\em Escherischia coli} (on the left) and
{\em Saccharomyces cerevisiae S288c} (on the right) --- a zoom on the first 30 sizes of groups sorted in descending order}
\end{figure}

The lists of identical genes are presented in Tables~\ref{TabBorr}, \ref{TabEColi} and~\ref{TabSac} (the genes are identified with labels originated from GeneBank files). Each table shows groups of genes sorted from the largest groups to the smallest ones. Respectively to the data in the tables two histograms with sorted group sizes for {\em Escherischia coli} and {\em Saccharomyces cerevisiae S288c} are presented in Figure~\ref{FigHist} (the histogram for {\em Borrelia burgdorferi} is omitted due to its simplicity --- just two groups which consist of only two genes appear in this case).

Figure~\ref{FigHist} shows how often the same sequences are repeated in genomes. For example, in the case of {\em Escherischia coli} the two mostly repeated sequence with the rank numbers one and two appear in 5 different genes each, the third sequence appears 4 times, etc.

The series with sizes of the groups of identical genes can be interpreted as probabilities of gene appearance in the genome.
To obtain this, for every unique sequence we assign its group size, that is the number of appearances in the genome. Then, the group sizes have to be normalized, because probabilities have to sum up to 1. Such probability of appearance of unique sequences is of hyperbolic type in the sense of definition $ p_{i}\geq i^{-\alpha}$, that is, one can show that the inequality is satisfied for each of the values in the series respectively for: {\em Borrelia burgdorferi} --- for, e.g.,  $\alpha = 9.2785$, {\em Escherischia coli}  --- for, e.g.,  $\alpha = 9.9426$, and {\em Saccharomyces cerevisiae S288c} --- for, e.g.,  $\alpha = 10.528$.

It is necessary to mention here, that due to the normalization of a series of nonzero values the inequality $p_{i}\geq i^{-\alpha}$ can never be satisfied for $i=1$ because $1^{-\alpha}=1$ regardless of $\alpha$ (simply, the probability of any option is never equal~1 if there is more that one random option to chose and all of them have a non-zero chance to be selected). Therefore, to show the hyperbolic nature of a single series, we needed to start with comparison of the second element $p_{i=2}$ and $2^{-\alpha}$, that is, $p_{i=1}$ is omitted. For clarity, we did also another calculations where the first element $p_{i=1}$ is compared with $(i+1)^{-\alpha}$ and so on, and obtained values of $\alpha$ are the same as in the former method.

Graphs of sizes of gene groups from Figure~\ref{FigHist} and the observation that probabilities of appearance of unique sequences are of hyperbolic type suggest that the realization of the theorem and that entropy loss is possible. Similarly, as in the case of languages the basic structure of genomes remains stabile.

\begin{table}
\centering
\caption{Groups of identical genes in the genome of {\em Borrelia burgdorferi}; every group starts with a group size, the length of a unique sequence which is written in brackets and the list of gene labels from GeneBank}
\label{TabBorr}
{\tiny
\centering
\begin{tabular}{cl}
  \hline
  2	(204) & gi|387827798|ref|NC$\_$017424.1|:15230-15433 B. burgd. N40 plasmid N40$\_$cp32-10\\
    & gi|387827867|ref|NC$\_$017402.1|:15177-15380 B. burgd. N40 plasmid N40$\_$cp32-9\\
  2	(237) & gi|387827798|ref|NC$\_$017424.1|:8732-8968 B. burgd. N40 plasmid N40$\_$cp32-10\\
    & gi|387827867|ref|NC$\_$017402.1|:8730-8966 B. burgd. N40 plasmid N40$\_$cp32-9\\
  \hline
\end{tabular}
}
\end{table}

\begin{table}[ht]
\centering
\caption{Groups of identical genes in the genome of {\em Escherischia coli}; every group starts with a group size, the length of a unique sequence which is written in brackets and the list of gene labels from GeneBank}
\label{TabEColi}
{\tiny
\centering
\begin{tabular}{cl}
  \hline
5	(378) & gi|387604868|ref|NC$\_$017627.1|:27400-27777 E. coli 042 plasmid pAA\\
& gi|387604868|ref|NC$\_$017627.1|:c40646-40269 E. coli 042 plasmid pAA\\
& gi|387605479|ref|NC$\_$017626.1|:c1601255-1600878 E. coli 042, complete genome\\
& gi|387605479|ref|NC$\_$017626.1|:c20942-20565 E. coli 042, complete genome\\
& gi|387605479|ref|NC$\_$017626.1|:c2746558-2746181 E. coli 042, complete genome\\
5	(522) & gi|387605479|ref|NC$\_$017626.1|:4574451-4574972 E. coli 042, complete genome\\
& gi|387605479|ref|NC$\_$017626.1|:565707-566228 E. coli 042, complete genome\\
& gi|387605479|ref|NC$\_$017626.1|:c1238313-1237792 E. coli 042, complete genome\\
& gi|387605479|ref|NC$\_$017626.1|:c1633110-1632589 E. coli 042, complete genome\\
& gi|387605479|ref|NC$\_$017626.1|:c7035-6514 E. coli 042, complete genome\\
4	(723) & gi|387605479|ref|NC$\_$017626.1|:2969552-2970274 E. coli 042, complete genome\\
& gi|387605479|ref|NC$\_$017626.1|:4575077-4575799 E. coli 042, complete genome\\
& gi|387605479|ref|NC$\_$017626.1|:c1632484-1631762 E. coli 042, complete genome\\
& gi|387605479|ref|NC$\_$017626.1|:c6409-5687 E. coli 042, complete genome\\
3	(276) & gi|387605479|ref|NC$\_$017626.1|:c4352916-4352641 E. coli 042, complete genome\\
& gi|387605479|ref|NC$\_$017626.1|:c4377124-4376849 E. coli 042, complete genome\\
& gi|387605479|ref|NC$\_$017626.1|:c665916-665641 E. coli 042, complete genome\\
3	(504) & gi|387605479|ref|NC$\_$017626.1|:c4352722-4352219 E. coli 042, complete genome\\
& gi|387605479|ref|NC$\_$017626.1|:c4376930-4376427 E. coli 042, complete genome\\
& gi|387605479|ref|NC$\_$017626.1|:c665722-665219 E. coli 042, complete genome\\
2	(276) & gi|387604868|ref|NC$\_$017627.1|:27080-27355 E. coli 042 plasmid pAA\\
& gi|387605479|ref|NC$\_$017626.1|:c4946969-4946694 E. coli 042, complete genome\\
2	(276) & gi|387604868|ref|NC$\_$017627.1|:c40966-40691 E. coli 042 plasmid pAA\\
& gi|387605479|ref|NC$\_$017626.1|:c2746878-2746603 E. coli 042, complete genome\\
2	(276) & gi|387605479|ref|NC$\_$017626.1|:c1601575-1601300 E. coli 042, complete genome\\
& gi|387605479|ref|NC$\_$017626.1|:c21262-20987 E. coli 042, complete genome\\
2	(858) & gi|387604868|ref|NC$\_$017627.1|:13-870 E. coli 042 plasmid pAA\\
& gi|387604868|ref|NC$\_$017627.1|:c108174-107317 E. coli 042 plasmid pAA\\
2	(258) & gi|387604868|ref|NC$\_$017627.1|:112802-113059 E. coli 042 plasmid pAA\\
& gi|387604868|ref|NC$\_$017627.1|:c108731-108474 E. coli 042 plasmid pAA\\
2	(177) & gi|387605479|ref|NC$\_$017626.1|:1409458-1409634 E. coli 042, complete genome\\
& gi|387605479|ref|NC$\_$017626.1|:c2263053-2262877 E. coli 042, complete genome\\
2	(492) & gi|387605479|ref|NC$\_$017626.1|:3438290-3438781 E. coli 042, complete genome\\
& gi|387605479|ref|NC$\_$017626.1|:5148233-5148724 E. coli 042, complete genome\\
2	(183) & gi|387605479|ref|NC$\_$017626.1|:1409283-1409465 E. coli 042, complete genome\\
& gi|387605479|ref|NC$\_$017626.1|:c2263228-2263046 E. coli 042, complete genome\\
2	(891) & gi|387604868|ref|NC$\_$017627.1|:33619-34509 E. coli 042 plasmid pAA\\
& gi|387604868|ref|NC$\_$017627.1|:c110418-109528 E. coli 042 plasmid pAA\\
2	(408) & gi|387605479|ref|NC$\_$017626.1|:c2327742-2327335 E. coli 042, complete genome\\
& gi|387605479|ref|NC$\_$017626.1|:c5136565-5136158 E. coli 042, complete genome\\
2	(327) & gi|387604868|ref|NC$\_$017627.1|:33296-33622 E. coli 042 plasmid pAA\\
& gi|387604868|ref|NC$\_$017627.1|:c21490-21164 E. coli 042 plasmid pAA\\
2	(108) & gi|387605479|ref|NC$\_$017626.1|:c4080133-4080026 E. coli 042, complete genome\\
& gi|387605479|ref|NC$\_$017626.1|:c4081099-4080992 E. coli 042, complete genome\\
2	(156) & gi|387605479|ref|NC$\_$017626.1|:1411502-1411657 E. coli 042, complete genome\\
& gi|387605479|ref|NC$\_$017626.1|:c2260557-2260402 E. coli 042, complete genome\\
  \hline
\end{tabular}
}
\end{table}

\newpage

\begin{table}[ht]
\centering
\caption{Groups of identical genes in the genome of {\em Saccharomyces cerevisiae S288c}; every group starts with a group size, the length of a unique sequence which is written in brackets and the list of gene labels from GeneBank}
\label{TabSac}
{\tiny
\centering
\begin{tabular}{cl}
  \hline
4	(132) & gi|330443681|ref|NC$\_$001144.5|:c468958-468827 S. c. S288c chr. XII\\
&gi|330443681|ref|NC$\_$001144.5|:c472610-472479 S. c. S288c chr. XII\\
&gi|330443681|ref|NC$\_$001144.5|:c482190-482059 S. c. S288c chr. XII\\
&gi|330443681|ref|NC$\_$001144.5|:c485842-485711 S. c. S288c chr. XII\\
4	(1323)&gi|330443520|ref|NC$\_$001136.10|:1206997-1208319 S. c. S288c chr. IV\\
&gi|330443531|ref|NC$\_$001137.3|:c449024-447702 S. c. S288c chr. V\\
&gi|330443681|ref|NC$\_$001144.5|:c481601-480279 S. c. S288c chr. XII\\
&gi|330443753|ref|NC$\_$001148.4|:56748-58070 S. c. S288c chr. XVI\\
4	(1089)&gi|330443681|ref|NC$\_$001144.5|:c470405-469317 S. c. S288c chr. XII\\
&gi|330443681|ref|NC$\_$001144.5|:c474057-472969 S. c. S288c chr. XII\\
&gi|330443681|ref|NC$\_$001144.5|:c483637-482549 S. c. S288c chr. XII\\
&gi|330443681|ref|NC$\_$001144.5|:c487289-486201 S. c. S288c chr. XII\\
3	(1323)&gi|330443638|ref|NC$\_$001142.9|:472760-474082 S. c. S288c chr. X\\
&gi|330443681|ref|NC$\_$001144.5|:593438-594760 S. c. S288c chr. XII\\
&gi|330443688|ref|NC$\_$001145.3|:196628-197950 S. c. S288c chr. XIII\\
3	(345)&gi|330443681|ref|NC$\_$001144.5|:472113-472457 S. c. S288c chr. XII\\
&gi|330443681|ref|NC$\_$001144.5|:485345-485689 S. c. S288c chr. XII\\
&gi|330443681|ref|NC$\_$001144.5|:488997-489341 S. c. S288c chr. XII\\
3	(5391)&gi|330443520|ref|NC$\_$001136.10|:1526321-1531711 S. c. S288c chr. IV\\
&gi|330443681|ref|NC$\_$001144.5|:1072508-1077898 S. c. S288c chr. XII\\
&gi|330443743|ref|NC$\_$001147.6|:1085473-1090863 S. c. S288c chr. XV\\
2	(528)&gi|330443489|ref|NC$\_$001135.5|:13282-13809 S. c. S288c chr. III\\
&gi|330443489|ref|NC$\_$001135.5|:200442-200969 S. c. S288c chr. III\\
2	(186)&gi|330443590|ref|NC$\_$001140.6|:c212720-212535 S. c. S288c chr. VIII\\
&gi|330443590|ref|NC$\_$001140.6|:c214718-214533 S. c. S288c chr. VIII\\
2	(1314)&gi|330443743|ref|NC$\_$001147.6|:1080276-1081589 S. c. S288c chr. XV\\
&gi|330443753|ref|NC$\_$001148.4|:c10870-9557 S. c. S288c chr. XVI\\
2	(363)&gi|330443681|ref|NC$\_$001144.5|:c13445-13083 S. c. S288c chr. XII\\
&gi|330443715|ref|NC$\_$001146.8|:781918-782280 S. c. S288c chr. XIV\\
2	(363)&gi|330443595|ref|NC$\_$001141.2|:c9155-8793 S. c. S288c chr. IX\\
&gi|330443638|ref|NC$\_$001142.9|:c9138-8776 S. c. S288c chr. X\\
2	(1323)&gi|330443391|ref|NC$\_$001133.9|:c165866-164544 S. c. S288c chr. I\\
&gi|330443753|ref|NC$\_$001148.4|:c810269-808947 S. c. S288c chr. XVI\\
2	(1323)&gi|330443531|ref|NC$\_$001137.3|:c498124-496802 S. c. S288c chr. V\\
&gi|330443688|ref|NC$\_$001145.3|:c378325-377003 S. c. S288c chr. XIII\\
2	(1323)&gi|330443743|ref|NC$\_$001147.6|:595112-596434 S. c. S288c chr. XV\\
&gi|330443753|ref|NC$\_$001148.4|:844709-846031 S. c. S288c chr. XVI\\
2	(1323)&gi|330443520|ref|NC$\_$001136.10|:1096064-1097386 S. c. S288c chr. IV\\
&gi|330443578|ref|NC$\_$001139.9|:c823015-821693 S. c. S288c chr. VII\\
2	(5313)&gi|330443543|ref|NC$\_$001138.5|:138204-139496,139498-143517 S. c. S288c chr. VI\\
&gi|330443578|ref|NC$\_$001139.9|:811738-813030,813032-817051 S. c. S288c chr. VII\\
2	(1317)&gi|330443543|ref|NC$\_$001138.5|:138204-139520 S. c. S288c chr. VI\\
&gi|330443578|ref|NC$\_$001139.9|:811738-813054 S. c. S288c chr. VII\\
2	(5580)&gi|330443578|ref|NC$\_$001139.9|:1084864-1084882,1085031-1090591 S. c. S288c chr. VII\\
&gi|330443753|ref|NC$\_$001148.4|:c6007-5989,c5840-280 S. c. S288c chr. XVI\\
2	(651)&gi|330443531|ref|NC$\_$001137.3|:c5114-4602,c4322-4185 S. c. S288c chr. V\\
&gi|330443681|ref|NC$\_$001144.5|:1066572-1067084,1067364-1067501 S. c. S288c chr. XII\\
2	(1770)&gi|330443595|ref|NC$\_$001141.2|:c18553-16784 S. c. S288c chr. IX\\
&gi|330443638|ref|NC$\_$001142.9|:c18536-16767 S. c. S288c chr. X\\
2	(495)&gi|330443743|ref|NC$\_$001147.6|:1082718-1083212 S. c. S288c chr. XV\\
&gi|330443753|ref|NC$\_$001148.4|:c8427-7933 S. c. S288c chr. XVI\\
2	(633)&gi|330443489|ref|NC$\_$001135.5|:c13018-12386 S. c. S288c chr. III\\
&gi|330443489|ref|NC$\_$001135.5|:c200178-199546 S. c. S288c chr. III\\
2	(1140)&gi|330443482|ref|NC$\_$001134.8|:c811479-810340 S. c. S288c chr. II\\
&gi|330443688|ref|NC$\_$001145.3|:7244-8383 S. c. S288c chr. XIII\\
  \hline
\end{tabular}
}
\end{table}

\subsection{Part III -- statistics for replicated fragments}

One can see that the numbers of identical genes in a genome are rather small mostly due to the fact, that gene sizes differ to each other in many cases. In the third part of the experimental research we evaluate numbers of identical fragments of genes in the way where the $i$-th symbol of one gene sequence being compared is not necessarily aligned against the $i$-th symbol of the other. Clearly, the size of compared fragment is constant, however, the fragment starting position in the gene may vary. For example, in the case of {\em Borrelia burgdorferi} the size of the shortest gene equals 120 symbols whereas the size of the longest one -- 6501. Assuming that the fragment size equals 120, there is just one fragment in the shortest gene and 6381 fragments in the longest one: a fragment from the first symbol to the 120-th one, from the second to 121-st, from the third to 122-nd, and so on. This way, for the fragment size equal~120 all of the genes are compared to each other many times, that is, each gene can be compared as many times as the number of fragments can be selected inside it.

\begin{figure*}[ht]
\centering
\includegraphics[width=4.9cm]{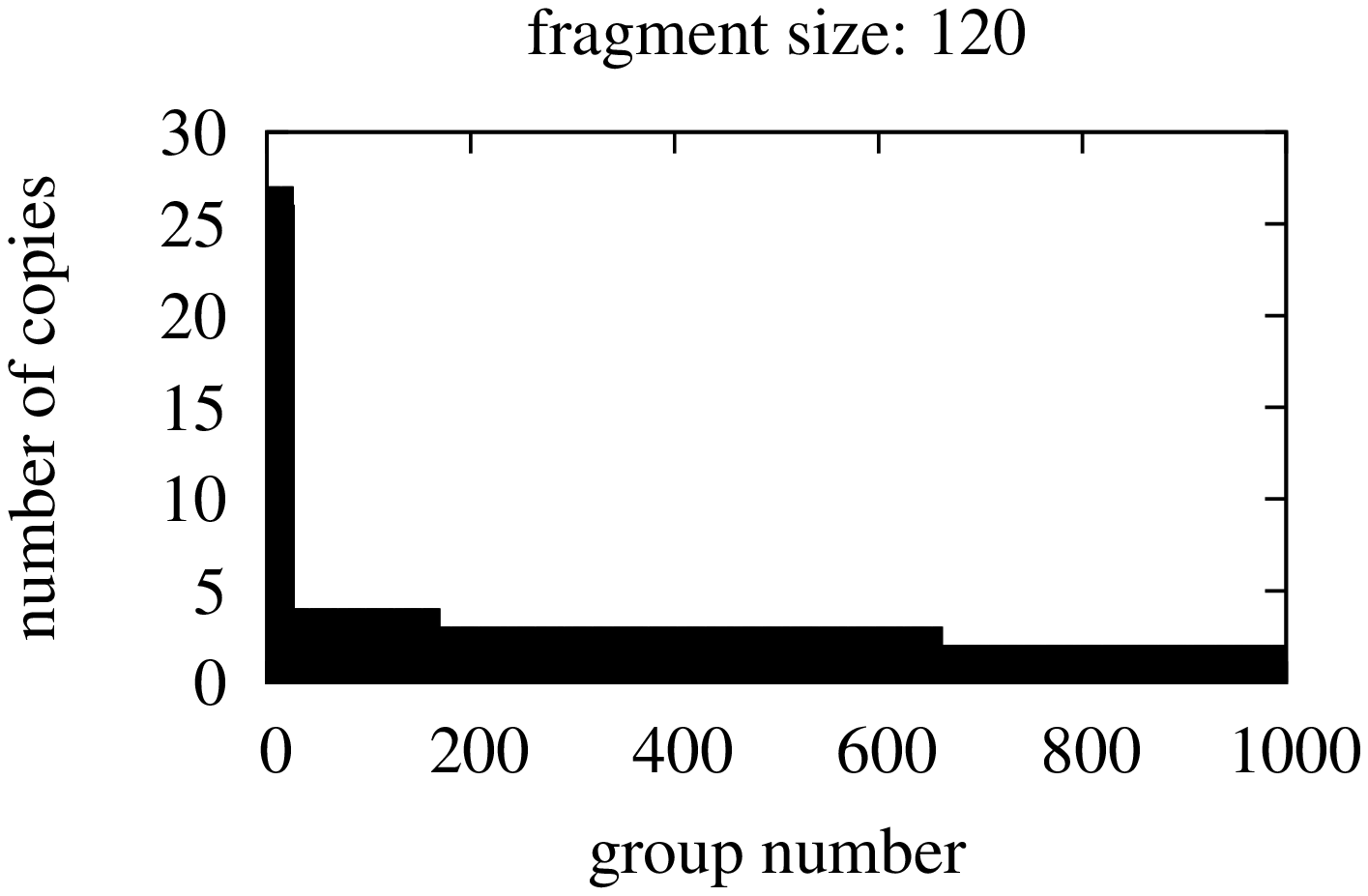}
\includegraphics[width=6.2cm]{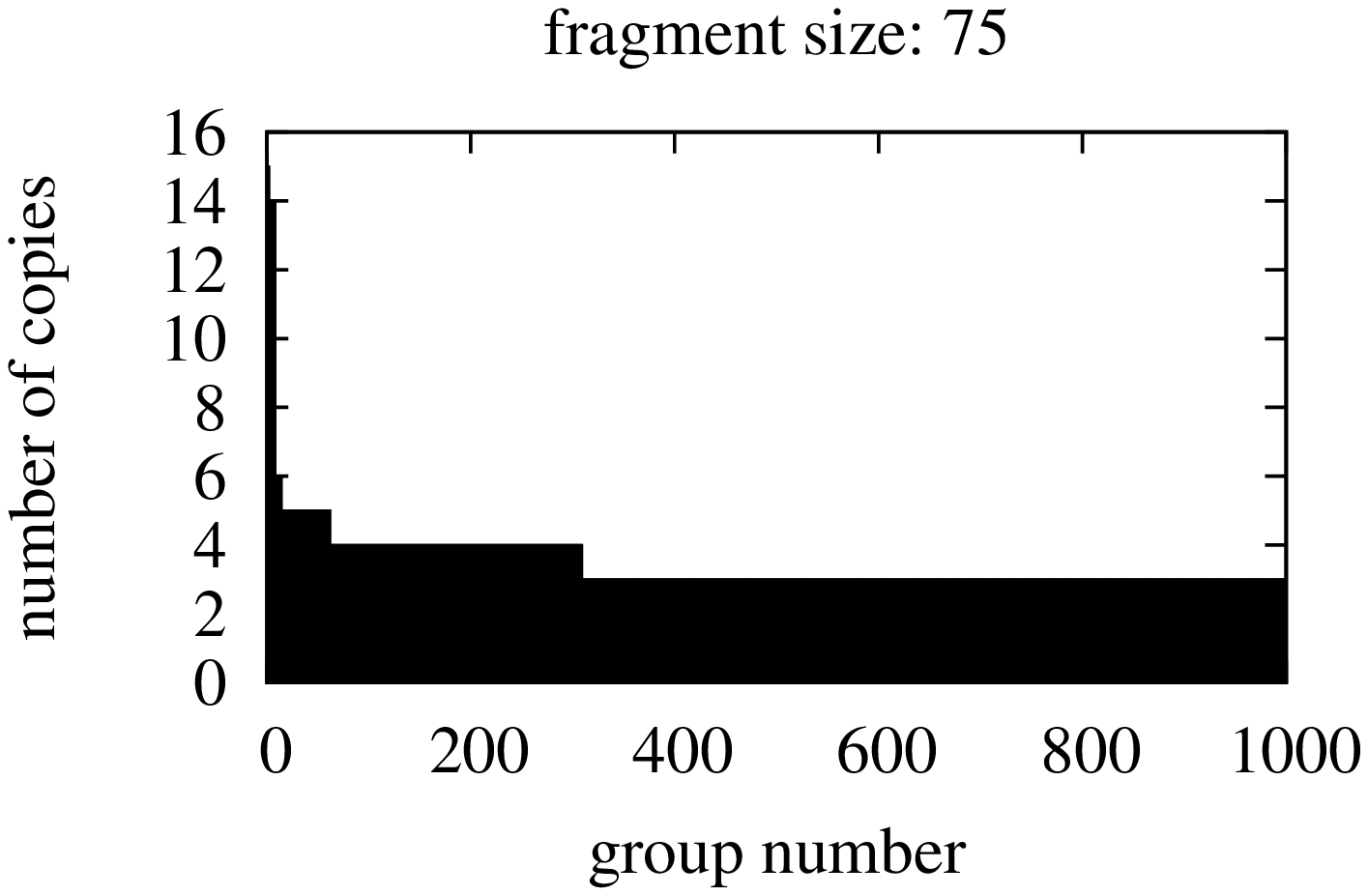}
\includegraphics[width=4.9cm]{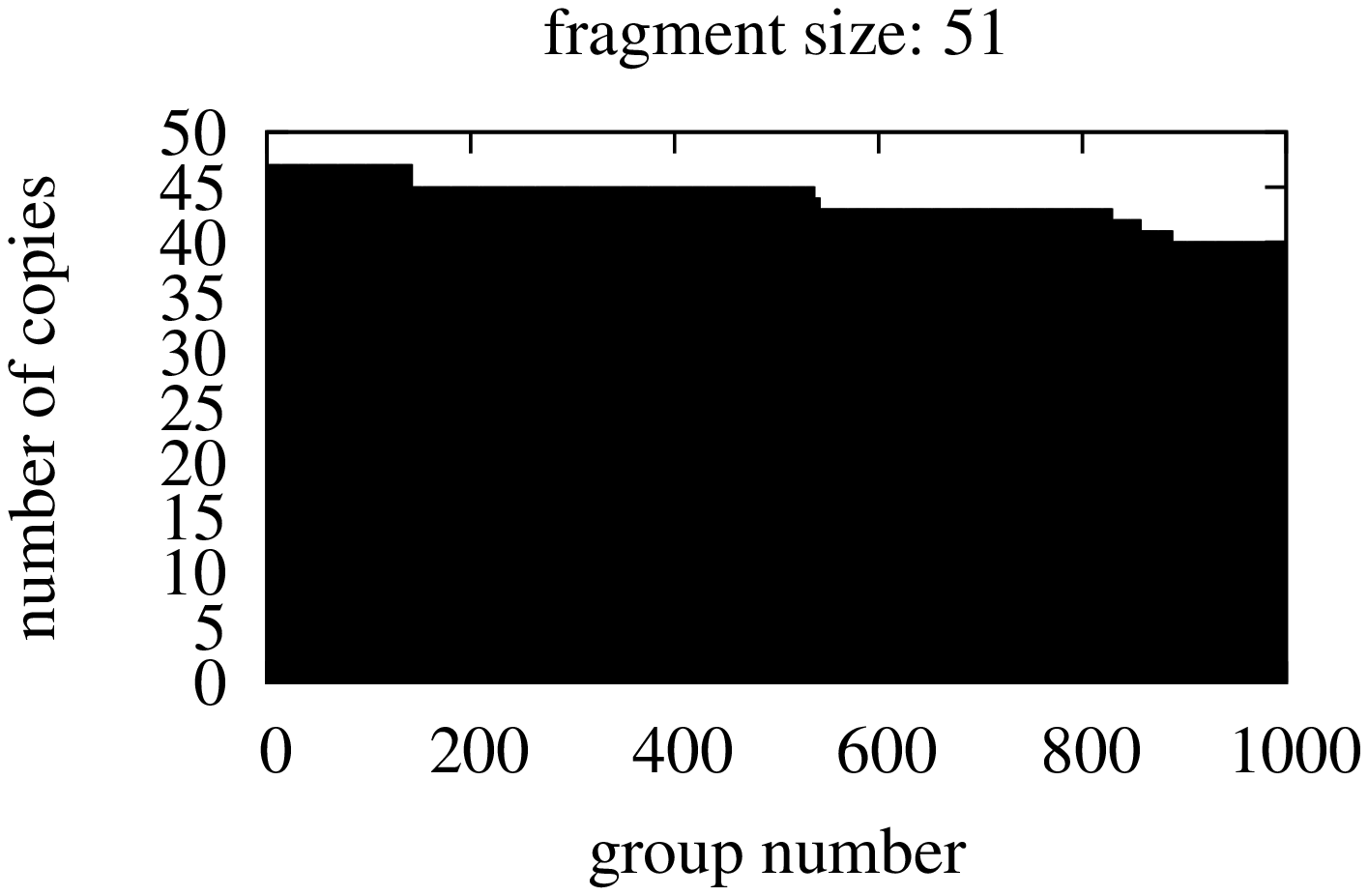}
\caption{\label{FHist} Graph of sizes of the groups of identical gene fragments in {\em Borrelia burgdorferi} (left), {\em Escherischia coli} (center) and {\em Saccharomyces cerevisiae S288c} (right) --- a zoom on the first 1000 sizes of groups sorted in descending order}
\end{figure*}

In this part we started from searching for identical fragments of size equal the size of the shortest gene in the genome, that is, fragments of size 120 for {\em Borrelia burgdorferi}, 75 -- for {\em Escherischia coli}, and 51 for {\em Saccharomyces cerevisiae S288c}. The figures with sizes of groups of identical fragments sorted in descending order are presented in Figure~\ref{FHist}.

In spite of the fact that in this comparisons the fragment starting position in compared genes may vary, both graphs in Figure~\ref{FigHist} and graphs in Figure~\ref{FHist} present similar "staircase" type. This confirms the stability and flexibility of  gene structures.

\begin{figure}[ht!]
\centering
\includegraphics[width=8cm]{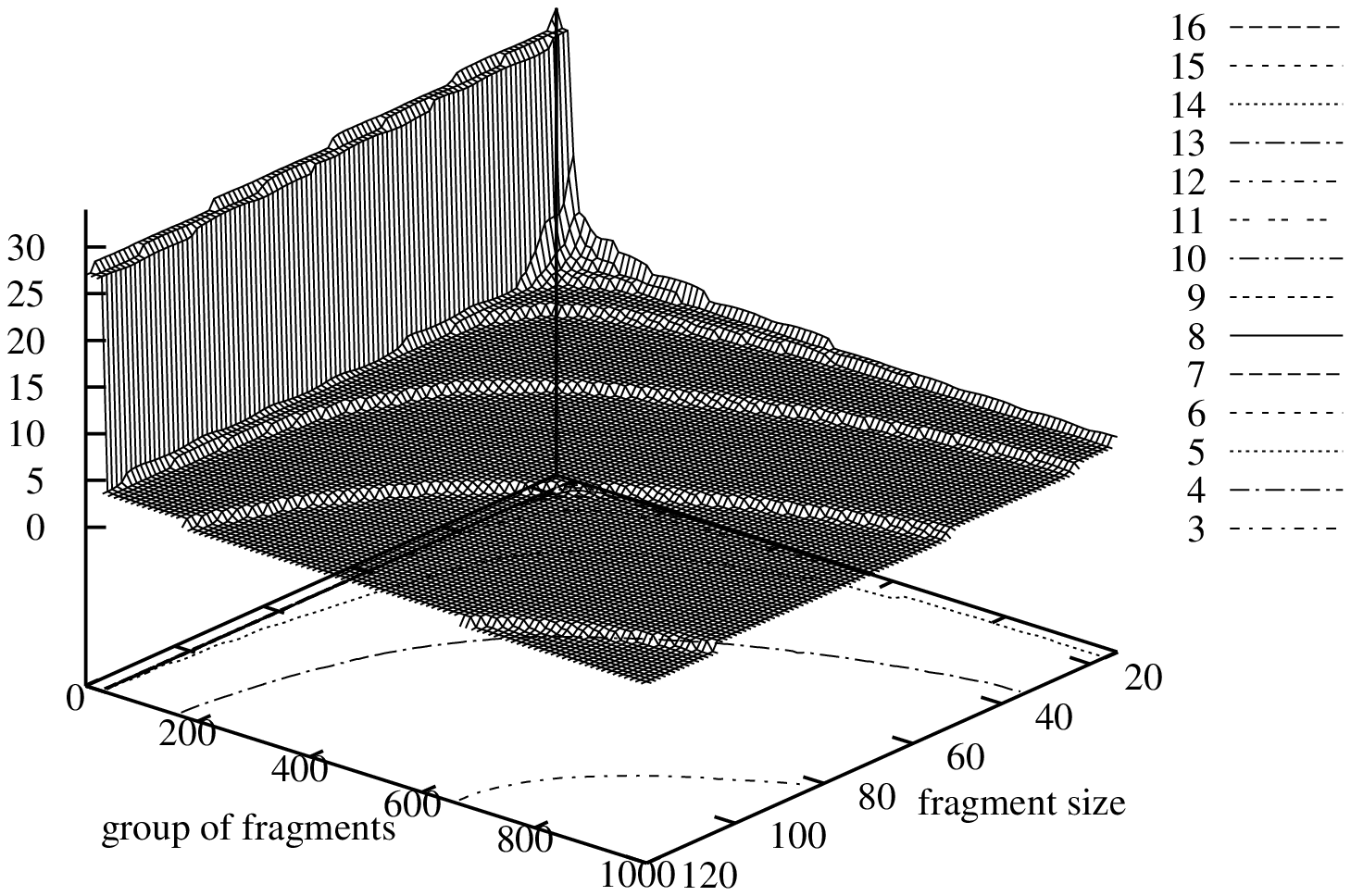}
\includegraphics[width=8cm]{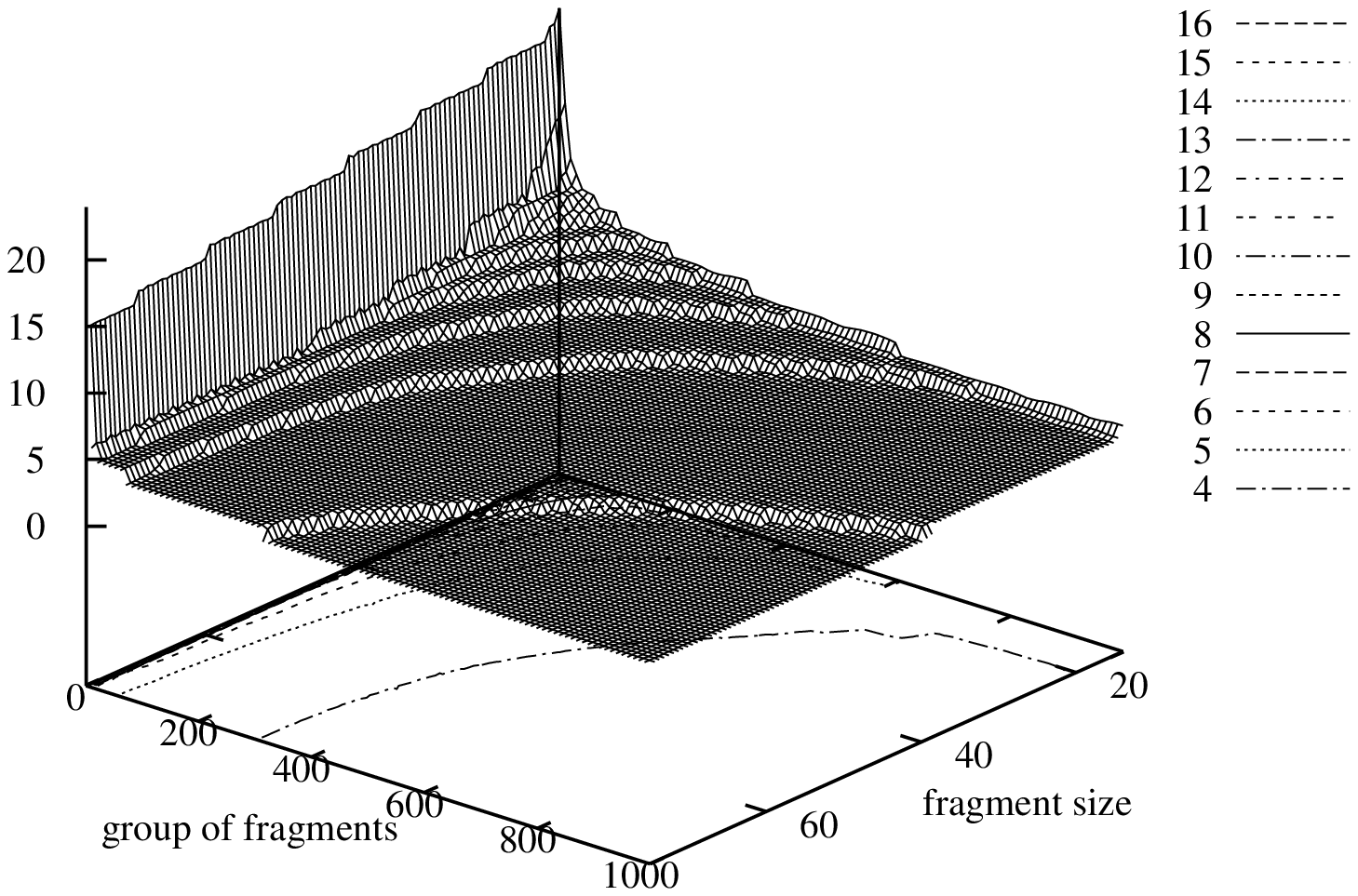}
\includegraphics[width=8cm]{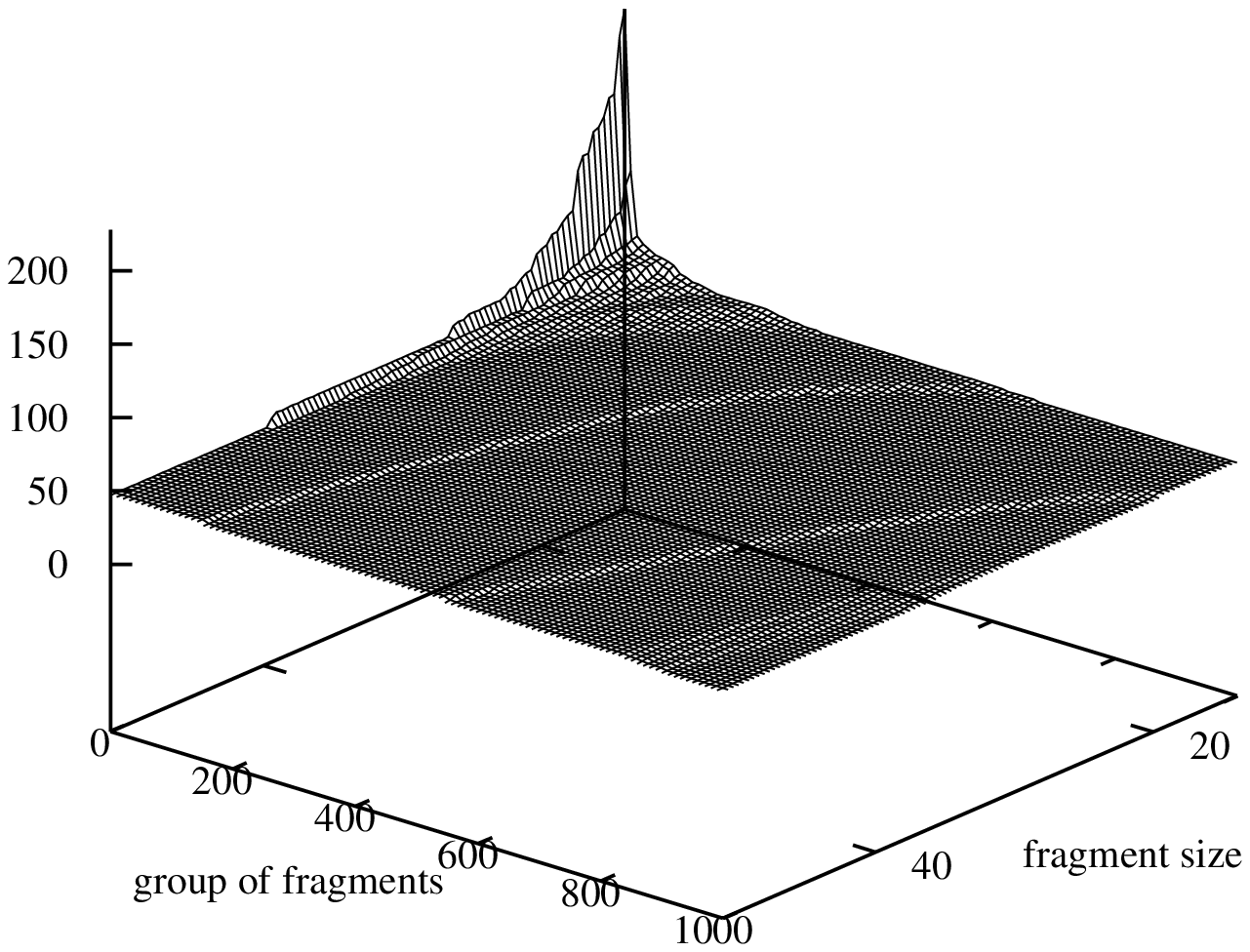}
\caption{\label{FH2} Graph of sizes of the groups of identical gene fragments in {\em Borrelia burgdorferi} (the left graph --- fragment sizes vary from 120 to 14), {\em Escherischia coli} (the right graph --- fragment sizes vary from 75 to 14) and {\em Saccharomyces cerevisiae S288c} (the bottom graph --- fragment sizes vary from 51 to 14)}
\end{figure}

It would be interesting to know how the characteristic of identical fragment numbers in a gene varies respectively to different sizes of the fragment size. Therefore, for each of the three genomes we evaluated numbers of identical fragments equal or smaller that the size of the shortest gene. For {\em Borrelia burgdorferi} the fragment size varied from 120 to 14, for {\em Escherischia coli} -- from 75 to 14, and for {\em Saccharomyces cerevisiae S288c} -- from 51 to 14. In every case the lower limit for the fragment length was set to 14 because the probability of fragment appearance in the gene grow rapidly for shorter fragments due to the limited number of permutations of four symbols in such a short sequence. Aggregated graphs with histograms of sizes of groups of identical fragments sorted in descending order are depicted in Figure~\ref{FH2}. The three graphs show a series of histograms obtained for each of the possible fragment sizes for the three genomes. The histograms all together build a surface over the two-dimensional domain where one axis represents a group number in the list sorted by the group size and the other -- a fragment size.

These figures present two dimensional surfaces with very regular curves. The graph shape and its regularity confirm stability and flexibility of gene structures.

\section{Final remarks}\label{Concl}
In this paper we follow the statistical description of genes in three selected genomes aimed at finding arguments supporting thesis that the entropy loss in gene "language" is possible and the basic structure of genomes remain stabile. We show that histograms of gene length and word length are similar and both can be modeled by Asymmetric Inverse Gaussian distributions. Additionally, tails of distributions for gene lengths appear to be of the power type. We test selected genomes and describe the gene code statistics dealing with the number of repetitions and replicated gene fragments in genomes. Considering that distributions of hyperbolic type describe property of stability and flexibility, we show that histograms of identical genes interpreted as probabilities of gene appearance in the genome
are of hyperbolic type in the sense of definition $ p_{i}\geq i^{-\alpha}$, and show example values of $\alpha$ for each of the genomes. We show also regularity and similarity of histograms with numbers of replicated genes and replicated gene fragments. All the obtained results confirm our thesis.

\begin{acknowledgements}
One of the authors (K. L-W) would like to thank prof. Franco Ferrari for discussions.
\end{acknowledgements}

\bibliography{KLWKTKT} 
\end{document}